\newif\ifllncs\llncsfalse
\newif\ifanon\anontrue
\newif\ifsub\subfalse

\ifllncs
\documentclass{llncs}
\else
\documentclass[11pt,letter]{article}
\fi

\usepackage{breakcites}
\usepackage[utf8]{inputenc}
\usepackage[T1]{fontenc}
\ifllncs
\else
\usepackage{times}
\usepackage{fullpage}
\fi

\usepackage{mdframed}
\usepackage{amsmath}
\usepackage{relsize}
\usepackage{amsfonts}
\usepackage{amssymb}
\usepackage[dvipsnames]{xcolor}
\usepackage{graphicx}
\usepackage{braket}
\usepackage{url}
\usepackage{bm}
\usepackage{tikz}
\usetikzlibrary{shapes,arrows,positioning,calc}
\usepackage[mathscr]{euscript}
\allowdisplaybreaks

\usepackage{stmaryrd}

\usepackage{soul}
\usepackage{multirow}
\definecolor{DarkBlue}{RGB}{0,0,150}
\definecolor{NotSoDarkBlue}{RGB}{15,15,210}
\definecolor{DarkRed}{RGB}{150,0,0}
\definecolor{DarkGreen}{RGB}{0,100,0}
\usepackage[pdfstartview=FitH,pdfpagemode=UseNone,colorlinks,linkcolor=DarkRed,filecolor=blue,citecolor=DarkRed,urlcolor=DarkRed,pagebackref,breaklinks]{hyperref}
\usepackage[ruled, algosection]{algorithm2e}
\usepackage{cleveref}

\newcommand{\poly}{\mathsf{poly}}

\usepackage{amsthm}
\newtheorem{theorem}{Theorem}

\newtheorem{conjecture}{Conjecture}
\newtheorem{lemma}[theorem]{Lemma}
\newtheorem{corollary}[theorem]{Corollary}
\newtheorem{definition}[theorem]{Definition}

\newtheorem*{theorem*}{Theorem}
\numberwithin{theorem}{section}
\numberwithin{conjecture}{section}
\numberwithin{problem}{section}

 \newtheorem*{conjecture*}{Conjecture}

\newmdtheoremenv[backgroundcolor=gray!10,
                 linewidth=0pt,
                 innerleftmargin=16pt,
                 innerrightmargin=16pt,
                 innertopmargin=6pt,
                 innerbottommargin=6pt,
            splitbottomskip=4pt]{protocol}[prot]{Game}

\def\matV{\mathbf{V}}
\def\matG{\mathbf{G}}
\def\matP{\mathbf{P}}
\def\matS{\mathbf{S}}

\newif\ifnotes
\ifsub
\notesfalse
\else
\notestrue
\fi

\title{Improving Algorithmic Efficiency using Cryptography}

\ifsub
\author{Anonymous Submission to FOCS 2025}
\else
\author{Vinod Vaikuntanathan\thanks{Research supported in part by DARPA under Agreement Number HR00112020023, NSF CNS-2154149, a Simons Investigator
award and a Thornton Family Faculty Research Innovation Fellowship from MIT.}\\MIT CSAIL \and Or Zamir\thanks{Research supported in part by the Israel Science Foundation, Grant No. 1593/24, and in part by the Blavatnik foundation. } \\Tel Aviv University}
\fi 
\date{}

\def\matA{\mathbf{A}}

\def\vecs{\mathbf{s}}

\def\vecv{\mathbf{v}}

\def\veca{\mathbf{a}}

\begin{document}
\maketitle


\def\matM{\mathbf{M}}
\def\matA{\mathbf{A}}
\def\matB{\mathbf{B}}
\def\matE{\mathbf{E}}
\def\bbF{\mathbb{F}}

\begin{abstract}
Cryptographic primitives have been used for various non-cryptographic objectives, such as eliminating or reducing randomness and interaction. We show how to use cryptography to \emph{improve the time complexity} of solving computational problems. Specifically, we show that under standard cryptographic assumptions, we can design algorithms that are asymptotically faster than existing ones while maintaining correctness.

As a concrete demonstration, we construct a distribution of \emph{trapdoored matrices} with the following properties: (a) computationally bounded adversaries cannot distinguish a random matrix from one drawn from this distribution (under computational hardness assumptions), and (b) given a trapdoor, we can multiply such an $n\times n$ matrix with any vector in near-linear (in $n$) time. We provide constructions both over finite fields and over the reals.
This enables a broad speedup technique: any algorithm relying on a random matrix—such as those that use various notions of dimensionality reduction—can replace it with a matrix from our distribution, achieving computational speedups while preserving correctness.

Using these trapdoored matrices, we present the first \emph{uniform} reduction from worst-case to approximate and average-case matrix multiplication with optimal parameters (improving on Hirahara-Shimizu STOC 2025, albeit under computational assumptions), the first worst-case to average-case reductions for matrix inversion, solving a linear system, and computing a determinant, as well as a speedup of inference time in classification models.
\end{abstract}
\thispagestyle{empty}
\newpage 

\ifsub
\else
\tableofcontents
\pagenumbering{roman}
\newpage
\fi 

\pagenumbering{arabic}

\section{Introduction}
Cryptographic techniques have long played a foundational role in securing communication, ensuring privacy, and enabling secure computation. However, beyond their traditional cryptographic applications, such techniques have also been leveraged to achieve non-cryptographic objectives. Notable examples include using cryptographic hardness assumptions or notions inspired by cryptography to reduce randomness in algorithms (e.g. pseudorandom generators starting from Blum-Micali~\cite{blummicali82,haastad1999pseudorandom,nisan1994hardness,impagliazzo1997p}), and  eliminate interaction in protocols (e.g. the celebrated Fiat-Shamir heuristic~\cite{fiat1986prove,DBLP:conf/stoc/CanettiCHLRRW19}). These applications illustrate how computational hardness assumptions can provide algorithmic benefits beyond the obvious goals of security and privacy.

In this work, we explore a novel use of cryptography: improving the {\em time complexity} of solving computational problems. Specifically, we demonstrate that under standard cryptographic assumptions, we can design algorithms that are {\em asymptotically faster} than existing ones while ensuring correctness on input instances generated by computationally bounded adversaries. We give a simple demonstration of this idea by leveraging structured randomness derived from cryptographic primitives to accelerate common computational tasks.

To illustrate this approach, we introduce the notion of \emph{trapdoored matrices}, a class of pseudorandom matrices that allow efficient multiplication when a trapdoor is available. We construct a distribution of such $n\times n$ matrices that satisfies the following key properties:
\begin{itemize}
\item\textbf{Indistinguishability:} A computationally bounded adversary cannot distinguish between a truly random matrix and one sampled from our distribution; and
\item\textbf{Efficient Multiplication:} Given a trapdoor, matrix-vector multiplication can be performed in near-linear $\tilde{O}(n)$ time. As an immediate consequence, matrix-matrix multiplication can be performed in near-quadratic $\tilde{O}(n^2)$ time.
\end{itemize}
We provide constructions of trapdoored matrices over both finite fields and over the reals. Over a finite field, we provide two constructions: one whose indistinguishability holds under the hardness of the well-known learning parity with noise problem (LPN)~\cite{DBLP:conf/crypto/BlumFKL93}; and a second based on the McEliece assumption, the basis of McEliece's cryptosystem from the 1970s~\cite{McEliece1978}. Over the reals, our construction relies on the pseudorandomness of the Kac random walk~\cite{kac1956foundations}; we introduce and motivate the relevant hardness assumption.

As a consequence, any algorithm that relies on a random matrix—such as those used in dimensionality reduction or sketching—can instead use a trapdoored matrix, achieving significant speedups due to the availability of efficient multiplication. Crucially, the cryptographic notion of indistinguishability enables a black-box reduction from the correctness of the modified algorithm to that of the original.
In particular, the reduction is independent of the specific properties of the random matrix required by the algorithm and its correctness proof. We remark that in fact, the hardness assumptions required for algorithmic correctness are milder than what one typically assumes in cryptography, e.g. assuming that LPN is hard for cubic-time (as opposed to any polynomial-time) algorithms suffices if the underlying algorithm runs in cubic time.
Moreover, we use these hardness assumptions in a \emph{conceptually weaker} manner than their standard use in cryptography: usually, it is assumed that a computationally-bounded yet possibly highly motivated player \emph{tries} to break the conjecture. 
In our case, the pseudorandomness stemming from the cryptographic conjectures is not meant to trick a dedicated adversary, but an {\em innocent} algorithm designed to solve a completely non-cryptographic problem.

\subsection{Applications of Random Matrices in Algorithm Design}
Random matrices play a fundamental role in modern algorithm design, particularly in settings where dimensionality reduction, efficient approximation, or sketching techniques are required. One of the most well-known applications is the \emph{Johnson-Lindenstrauss (JL) Lemma}, which states that a set of~$N$ points in a possibly high-dimensional Euclidean space can be embedded into a~$\Theta\left((\log N) / \varepsilon^2\right)$-dimensional space while approximately preserving pairwise Euclidean distances up to~$\left(1\pm \varepsilon\right)$ multiplicative error. The standard JL transform relies on mapping the set of points through a random projection matrix, often drawn from Gaussian or sub-Gaussian distributions~\cite{johnson1984extensions,DBLP:conf/stoc/IndykM98}.

Beyond the JL lemma, random matrices and the idea of randomized dimensionality reduction are used in various other algorithmic settings, including:
\begin{itemize}
\item
\textit{Sketching and Streaming:} Randomized sketching techniques, such as AMS, CountSketch, and various other streaming algorithms, enable efficient approximation of high-dimensional data with lower memory and computation costs~\cite{alon1996space,charikar2002finding}.
\item
\textit{Randomized Singular Value Decomposition (SVD):} Techniques such as those introduced by Halko, Martinsson, and Tropp~\cite{halko2011finding} use random projections to construct a low-rank approximation of a matrix. This is used to efficiently approximate dominant singular values and singular vectors, with applications in machine learning and numerical linear algebra.
\item
\textit{Subspace Embeddings:} Random matrices are used to construct oblivious subspace embeddings~\cite{sarlos2006improved,woodruff2014sketching}, which facilitate efficient least squares regression, low-rank approximation, and optimization problems by approximately preserving the geometry of a subspace.
\item
\textit{Spectral Sparsifiers:} Random sampling techniques, some of which involve random matrices, help construct spectral sparsifiers~\cite{spielman2011spectral}, which approximate the spectral properties of a graph while significantly reducing the number of edges. These sparsifiers are useful in solving linear systems, graph algorithms, and optimization.
\item
\textit{Error-Correcting Codes (ECCs):} Random matrices play a fundamental role in coding theory, particularly in the construction of random linear codes, where a randomly generated generator matrix is used to encode messages into redundant codewords for error correction~\cite{varshamov1957estimate}.
\item
\textit{Dimensionality Reduction Beyond JL:} The JL lemma is optimal in the sense that~$\Omega\left((\log N) / \varepsilon^2\right)$ dimensions are needed to preserve all pair-wise distances for \emph{any} Euclidean set of~$N$ points~\cite{larsen2017optimality}.
On the other hand, if additional assumptions are made regarding the set of points, a reduction to a lower dimension is sometimes possible.
For example, better bounds exist when the set of points is assumed to be of low \emph{intrinsic dimension} (e.g., doubling dimension) or when not all pair-wise distances are required to be preserved~\cite{indyk2007nearest,boutsidis2010random,cohen2015dimensionality,makarychev2019performance,narayanan2021randomized}.
\end{itemize}

In all of these applications, random matrices are integral to the algorithm: they are multiplied by vectors or other matrices, often becoming the computational bottleneck. Since a random $m$-by-$n$ matrix is dense and unstructured, even multiplying it by a single vector takes~$\Theta(mn)$ time.

To mitigate this bottleneck, researchers have sought alternative constructions that preserve the key properties of random matrices while allowing for more efficient computation. One prominent example is the extensive study of \emph{fast JL transforms}~(e.g., \cite{ailon2009fast,ailon2013almost,bamberger2021optimal,jain2022fast}), which provide fast, low-dimensional embeddings that preserve pairwise distances, similar to traditional Johnson-Lindenstrauss (JL) random projections, but with significantly reduced computational cost---culminating in constructions that run in near-linear time. Another notable example is the development of \emph{error-correcting codes with fast encoding}~\cite{gelfand1973complexity,DBLP:conf/stoc/Spielman95}.

However, these tailored constructions lack the full flexibility of truly random linear transformations. A fast transform designed to preserve pairwise distances may not necessarily maintain other structural properties, such as spectral norms or subspaces, making it unsuitable for the applications it was not specifically designed for.
Our core idea is to replace these randomly sampled matrices with structured, yet pseudorandom matrices---\emph{trapdoored matrices}---that retain the algorithmic utility of fully random matrices while allowing for faster multiplication. This enables a black-box transformation: \emph{any} algorithm that relies on random matrices can be accelerated without requiring new, problem-specific correctness proofs, making our approach broadly applicable across a wide range of applications.

\subsection{Our Results}\label{sec:defs}

We now present our results in more detail, starting with our notion of \emph{trapdoored matrices}. We then provide several constructions of trapdoored matrices over different fields and based on different cryptographic assumptions.

An efficiently samplable distribution $\mathcal{D}$ over pairs $(M,C)$ where $M$ is a matrix over~$\mathbb{F}^{n \times m}$ (where $\mathbb{F}$ is a field, either a finite field or the reals in this work)  and $C:\mathbb{F}^m \to \mathbb{F}^n$ is a circuit, is a trapdoored matrix distribution if:
\begin{itemize} 
    \item (Correctness) \;For \emph{every} vector~$v\in \mathbb{F}^n$, the circuit $C$ computes the matrix-vector multiplication~$$C(v)=Mv.$$
    \item (Efficiency) \;The circuit $C$ is small: it has size $\widetilde{O}(n+m)$.
    \item (Indistinguishability) \;A matrix $M$ drawn from $\mathcal{D}$ is computationally indistinguishable from a uniformly random matrix over $\mathbb{F}^{n\times m}$ (or from an appropriate finite distribution, e.g. a truncated and discretized Gaussian, if the field is the reals).
\end{itemize} 

In other words, the circuit $C$ is the trapdoor (associated to $M$) that allows for quick computation of the matrix-vector product with any given vector $v$. 

The main conceptual contribution of our work is the following simple yet powerful insight: computational indistinguishability ensures that any algorithm using random matrices can instead use matrices drawn from a trapdoored matrix distribution without affecting correctness or success probability. Any observable difference in performance would itself serve as a polynomial-time \emph{distinguisher} between the uniform and trapdoored matrix distributions, contradicting indistinguishability. Thus, we can replace matrix multiplications within the algorithm with the corresponding trapdoor circuits while preserving correctness, as these circuits produce exactly the same outputs -- just more efficiently.

We emphasize that correctness holds as long as the algorithm’s inputs are not chosen \emph{dependent on the trapdoor circuits}~$C$. However, they may depend on the matrices themselves, and in particular on previous computations involving them.

We also note that while we crucially rely on the full generality of indistinguishability to preserve correctness in a black-box manner, our reliance on hardness assumptions is actually \emph{weaker} than in most works in classical cryptography.
Typically, in cryptographic settings, a computationally bounded adversary is assumed to actively \emph{attempt} to break the assumption. In contrast, in our work, the potential distinguisher is merely a natural algorithm designed for an unrelated, non-cryptographic problem. As a result, it is even less likely that such a distinguisher exists for any of the hardness assumptions we employ in this paper.

\paragraph{Our Constructions of Trapdoored Matrices.}
We first observe that a construction of near-linear-time-efficient trapdoored matrices over \emph{square matrices} actually suffices to get one also for rectangular matrices or for matrix-matrix multiplication (rather than matrix-vector multiplication). 
This is because a rectangular matrix can be simply written as a concatenation of square ones, and matrix-matrix multiplication can be computed column-by-column. 
Furthermore, in some of our constructions we can also construct a near-linear time circuit computing matrix-vector multiplications with the matrix inverse~$M^{-1}$ and not only with the matrix~$M$ itself; this feature might be useful for certain applications.

Our first two constructions give near-linear-time-efficient trapdoored matrices for the uniform distribution of square matrices over a finite field~$\mathbb{F}$. Each of the two constructions is based on a different hardness assumption.
\begin{theorem*}[Section~\ref{sec:lpn}]
  Under the polynomial hardness of the learning parity with noise assumption over a field $\mathbb{F}$, there is a collection of trapdoored matrices for which, given the trapdoor, $n\times n$ matrix-vector multiplication can be performed in time $n^{1+\epsilon}$ for an arbitrarily small constant $\epsilon>0$. Assuming sub-exponential hardness, this can be done in time 
  $n\cdot \mathsf{poly}(\log n)$. 
\end{theorem*}

The learning parity with noise (LPN) assumption is a classical average-case hardness assumption originating from learning theory and coding theory~\cite{DBLP:conf/crypto/BlumFKL93,DBLP:conf/focs/Alekhnovich03}. For a detailed discussion of the assumption, we refer the reader to Section~\ref{sec:lpnassm}.

\begin{theorem*}[Section~\ref{sec:mcel}]
  Under the polynomial hardness of the McEliece assumption over a field $\mathbb{F}$, there is a collection of trapdoored matrices for which, given the trapdoor, $n\times n$ matrix-vector multiplication can be performed in time $n^{1+\epsilon}$ for an arbitrarily small constant $\epsilon>0$. Assuming sub-exponential hardness, this can be done in time $n\cdot \mathsf{poly}(\log n)$. 
\end{theorem*}

The McEliece assumption is a formalization of an assumption underlying the security of the McEliece public-key cryptosystem~\cite{McEliece1978} from the 1970s, also the basis of the NIST post-quantum cryptography submission ``Classic McEliece''~\cite{ClassicMcE}. For a detailed discussion of the assumption, we refer the reader to Section~\ref{sec:mceassm}. 

Our third construction is for distributions of real-valued matrices, such as those used in JL-like dimension reductions.
We say that a distribution~$\mathcal{D}$ over matrices is \textbf{\emph{Haar-invariant}} if the matrix products $O\cdot A$ and~$A\cdot O$, where~$A\sim \mathcal{D}$ and~$O$ is sampled from the Haar-measure over~$SO(n)$, are both distributed as~$\mathcal{D}$.
Informally, these are all distributions that are invariant to a random rotation of either the input or output space. 
An example for such a distribution~$\mathcal{D}$ is the distribution of matrices in which every coordinate is an i.i.d. normal Gaussian variable. 
We say that a distribution~$\mathcal{D}$ over \emph{symmetric} matrices is \textbf{\emph{symmetrically Haar-invariant}} if the matrix product $O\cdot A \cdot O^{-1}$, where~$A\sim \mathcal{D}$ and~$O$ is sampled from the Haar-measure over~$SO(n)$, is distributed as~$\mathcal{D}$.
Informally, these are all distributions that are invariant to conjugation by a random rotation of the space. 
An example of such a distribution is the Gaussian Orthogonal Ensemble (GOE).

\begin{theorem*}[Section~\ref{sec:reals}]
Let~$\mathcal{D}$ be \emph{any} distribution that is either Haar-invariant or symmetrically Haar-invariant.
  Under the psuedorandomness of Kac's random walks, there is a collection of trapdoored matrices over the distribution~$\mathcal{D}$ for which, given the trapdoor, $n\times n$ matrix-vector multiplication can be performed in time $n\cdot \mathsf{poly}(\log n)$. 
\end{theorem*}

For a detailed discussion of Kac's random walks and their plausible pseudorandomness, we refer the reader to Section~\ref{sec:kacassm}. A similar assumption in the context of finite fields was studied by Sotiraki~\cite{sotiraki} in the context of constructing fine-grained key-exchange protocols in the spirit of Merkle~\cite{DBLP:journals/cacm/Merkle78}.

\paragraph{A Bird's Eye View of Our Techniques.}
All our constructions start by identifying special classes of matrices for which matrix-vector multiplication can be performed in near linear time. Such matrices include low-rank matrices, sparse matrices, matrices with Vandermonde-like structure, and a product of a small number of elementary matrices. We then proceed to randomize these special matrices so as to hide their structure, but also preserve near-linear-time computability of matrix-vector products. Our LPN construction employs a combination of sparse and low-rank matrices; our McEliece construction (using Goppa codes) employs the fact that batch polynomial evaluation can be done fast; and our construction of trapdoored matrices over the reals uses Kac's random walk, i.e. products of a small number of elementary matrices. 

Beyond this starting point, our constructions require further ideas. For example, our basic LPN-based construction achieves $n^{3/2}$-time matrix-vector multiplication, worse than our eventual goal. We augment this basic construction with recursion to achieve near-linear-time computation. The same goes for our construction based on the McEliece assumption. Moving on to trapdoor matrices over the reals, our basic and natural construction samples trapdoored matrices for the uniform distribution over $SO(n)$, the special orthogonal group. We then augment this basic construction with a reduction showing how to use such a sampler to construct trapdoored matrices for any Haar-invariant distribution, in particular the Gaussian distribution.

\subsection{Motivating Applications}

\subsubsection{Faster Inference for Classification Models}
We highlight an application where fast transforms are not known, demonstrating the versatility of our approach.
It was recently shown that for the task of preserving the \emph{optimal clustering} of a point set, a random projection to a much lower dimension than the JL bound suffices.
Optimal bounds on the dimension were proven when parametrized by the number of clusters~\cite{makarychev2019performance} or by an intrinsic dimension of the point set~\cite{narayanan2021randomized}.
These results rely on intricate arguments beyond standard applications of the JL lemma.

These findings found a compelling application in \emph{machine learning}: when training a \emph{classifier} (i.e., a model that distinguishes between different \emph{classes} of inputs), the information often comes from the clustering structure of the data.
Thus, if we apply a transformation that preserves optimal clustering, we can reduce the dimensionality of the training set before running the training algorithm, significantly lowering the computational cost of training and reducing the number of parameters in the model.
At inference time, we would apply the same transformation to new inputs before classification.
Empirical results confirm the viability of this approach:~\cite{narayanan2021randomized} demonstrated that applying a random linear transformation to commonly studied datasets (such as the MNIST handwritten digits dataset and the ISOMAP face image dataset~\cite{tenenbaum2000global}) preserves the separation to classes even after significant dimensionality reduction.

While this method accelerates training, it does not improve the \emph{inference} time since the cost of applying the random linear transformation for dimensionality reduction is significant in comparison to the cost of running the neural network inference on the lower-dimensional data. 
By replacing the random linear transformation with a {trapdoored matrix}, we not only retain the training-time savings but also achieve \emph{fast inference} by reducing the cost of applying the initial transformation. (This could also reduce the training time, but there, the cost of the initial dimension-reduction of the data--- a rectangular matrix multiplication--- is often {\em not} the dominant cost.)

\subsubsection{Uniform Error-Correction for Matrix Multiplication}
Given a black-box algorithm that \emph{approximately} computes matrix multiplication \emph{for an average input}, can we design a nearly-as-efficient algorithm that computes matrix multiplication \emph{exactly for any input}?
Several variants of this question were extensively studied. 
Blum, Luby, and Rubinfeld~\cite{blum1990self} initially showed that given an algorithm that computes~$A\cdot B$ \emph{exactly} for a sufficiently large constant fraction of input pairs~$A,B\in \mathbb{F}_p^{n\times n}$, then we can also compute~$A\cdot B$ for \emph{any} input pair with high probability using almost the same amount of time.
Subsequent works~\cite{asadi2022worst,hirahara2023hardness} obtained the same result even when the given algorithm produces the correct multiplication only on a \emph{very small fraction} of input pairs.

Later works~\cite{gola2024matrix, hirahara2025error} the above was generalized to the case where the given algorithm only produces a matrix that is close to~$A\cdot B$ in Hamming distance, but is not guaranteed to exactly compute the multiplication for any input pair.
Recently, Hirahara and Shimizu~\cite{hirahara2025error} obtained a \emph{non-uniform} such reduction given the weakest possible multiplication algorithm over a field of prime order: Given an algorithm that on a random pair~$A,B\sim \mathbb{F}_p^{n\times n}$ of inputs returns a matrix whose relative Hamming distance from~$A\cdot B$ is at most~$1-\frac{1}{p}-\varepsilon$, for some~$\varepsilon>0$, they can compute~$A\cdot B$ exactly for any input pair~$A,B$ in a similar time. Note that an algorithm that simply returns a random matrix is in expectation only~$(1-\frac{1}{p})$-far from the correct multiplication, and they assume only a small advantage over that. 
On the other hand, their reduction is \emph{non-uniform}, meaning that they need some large pre-processing time that depends only on the matrix size~$n$.
Furthermore, all of the aforementioned results are based on highly sophisticated tools.

Using our construction of trapdoored matrices, we are able to obtain the same result using a \emph{uniform} reduction, affirmatively answering the main remaining problem in this line of work --- under a computational assumption. Furthermore, given a trapdoored matrix family, our reduction is extremely simple and works in any finite field.

\begin{theorem*}[\ref{thm:errcorrectMM}]
    Let~$\mathbb{F}_p$ be a finite field of order~$p$, and assume the existence of a trapdoored matrix family over~$\mathbb{F}_p$ and let~$\varepsilon>0$.
    Let~$\mathcal{M} : \mathbb{F}_p^{n\times n} \times \mathbb{F}_p^{n\times n} \rightarrow \mathbb{F}_p^{n\times n}$ be an algorithm running in time~$T(n)$ such that for every sufficiently large~$n$, 
    \[
    \mathbb{E}_{A,B\sim \mathbb{F}_p^{n\times n}} \Big[
    \text{dist}\left(A\cdot B, \ \mathcal{M}\left(A,B\right)\right) 
    \Big] \leq\left( 1-\frac{1}{p}-\varepsilon\right)n^2.
    \]
    Then, we can construct an algorithm that for \textbf{any} matrices $A,B\in \mathbb{F}_p^{n\times n}$ computes the multiplication~$A\cdot B$ \textbf{exactly} with high probability, in time~$\tilde{O}\left(T(n)\right)$.
\end{theorem*}

\subsubsection{Worst Case to Average Case Reductions for Other Linear Algebraic Operations}
Similar questions to those discussed above can be asked for other linear algebraic operations beyond matrix multiplication.
In Strassen's seminal paper~\cite{strassen1969gaussian}, which introduced the first improvement to the matrix multiplication exponent, he also showed that matrix inversion, determinant computation, and Gaussian elimination have the same computational complexity as matrix multiplication (up to lower-order terms).
To the best of our knowledge, no worst-case to average-case reductions are known for these problems, even when the average-case algorithm is guaranteed to produce a correct answer with probability at least~$99\%$.
In~\cite{blum1990self}, the authors described algorithms that solve these problems in the worst case, assuming access to average-case algorithms for \emph{all} of these problems (including, in particular, average-case matrix multiplication).
In contrast, the simplicity of our reductions arising from trapdoored matrices enables reductions that rely only on the average-case version of the same problem.

In Section~\ref{sec:lingalg}, we present reductions, using trapdoored matrix families such as those we construct in this paper, from the worst-case versions of matrix multiplication, matrix inversion, solving a system of linear equations with a unique solution and computing a determinant to the average-case version of each problem. 
As in the worst-case all of these problems are computationally equivalent, this implies that an average-case algorithm for \emph{any} of them would result in a worst-case algorithm for \emph{all} of them.

\subsection{Related Work}

Pseudorandomness has been used in the past to reduce \emph{space} in algorithmic tasks. 
A prominent example is \emph{algorithmic hash functions}: instead of using a fully random function, a succinctly represented \emph{pseudorandom} function often suffices for algorithmic applications-for instance, $k$-wise independent hash families can efficiently replace fully random functions in hash tables. Similarly, certain classes of expander graphs, such as those constructed as Cayley graphs, provide an explicit and succinct representation that eliminates the need to store a fully random graph in memory while preserving key properties.
Motivated by these examples, Goldreich, Goldwasser, and Nussboim~\cite{DBLP:journals/siamcomp/GoldreichGN10} initiated a broader line of research exploring how to implement large (pseudo-)random objects in a space-efficient manner, while also maintaining efficient access to these objects (such as applying a pseudo-random function on an input or accessing the adjacency lists of a pseudo-random graph). Their and subsequent works develop such implementations for pseudorandom functions and graphs, including both dense~\cite{DBLP:journals/siamcomp/GoldreichGN10} and sparse~\cite{DBLP:conf/approx/NaorN07} instances.

\paragraph{Note:} In a concurrent and independent work, Braverman and Newman~\cite{DBLP:journals/corr/abs-2502-13060} suggest and use a construction very similar to that of our Section~\ref{sec:lpn}, in a different context motivated by privacy-preserving delegation of computation. Our focus, on the other hand, is using these cryptographic primitives for \emph{non-cryptographic objectives}. Moreover, we present additional constructions based on McEliece-type assumptions and for real-valued matrices. 

\section{Trapdoored Matrices}

We start with a rigorous definition of our notion of \emph{trapdoored matrices}. We then provide several constructions of trapdoored matrices over different fields and based on different cryptographic assumptions. Let $\mathbb{F}$ be a field (either a finite field or the reals, in this work).

\begin{definition}[Trapdoored Matrices over~$\mathbb{F}^{n \times m}$]
    A distribution of \emph{trapdoored matrices} over~$\mathbb{F}^{n \times m}$ is an efficiently samplable distribution~$\mathcal{D}$ over pairs~$(M,C)$ in which the first coordinate is a matrix~$M\in \mathbb{F}^{n \times m}$ and the second coordinate is a \emph{circuit}~$C$, such that for \emph{every} vector~$v\in \mathbb{F}^n$ the circuit computes the matrix-vector multiplication~$C(v)=Mv$.
\end{definition}

To be considered good, we want the distribution $\mathcal{D}$ to have \emph{efficient multiplication} and \emph{indistinguishability}, which we define next.

\begin{definition}[Efficient Multiplication]
    For a time function~$T:\mathbb{N} \to \mathbb{N}$, we say that a trapdoored matrix distribution~$\mathcal{D}$ is~$T$-efficient if for every pair~$(M,C)$ sampled from~$\mathcal{D}$ and every vector~$v\in \mathbb{F}^n$, the circuit~$C$ can be applied to~$v$ in~$T(n)$ time.
\end{definition}

\begin{definition}[Indistinguishability]
    A trapdoored matrix distribution~$\mathcal{D}$ is  \emph{indistinguishable from the uniform distribution~$\mathcal{U}$} if every polynomial-time algorithm~$\mathcal{A}$ is unable to distinguish whether it is given a uniformly drawn random matrix from~$\mathcal{U}$ or a matrix~$M$ drawn using~$\mathcal{D}$, that is
    \[
    \left| \Pr\left[\mathcal{A}(M) = 1: M \sim \mathcal{U} \right]
    -
    \Pr\left[\mathcal{A}(M) = 1: (M,C) \sim \mathcal{D} \right]
    \right|
    \leq
    \mu(n),
    \]
    where~$\mu(n)$ is a function that vanishes quicker than any inverse polynomial in~$n$.
\end{definition}

This definition can be generalized in several ways: 
first, from matrix-vector multiplication to matrix-matrix multiplication; and secondly, by allowing for efficient multiplication on both the right and  the left; and thirdly, by allowing for efficient multiplication by both $M$ and $M^{-1}$. All our constructions can be easily extended to matrix-matrix multplication. Our LPN-based construction supports left and right multiplication. Our construction over the reals supports left and right multiplication, as well as multiplication by $M^{-1}$.

Yet another extension that one could conceive of would be to quickly output a pseudorandom matrix together with a property or class of properties, e.g. $M$ together with its determinant, rank, etc. We do not explore this extension further in this work.

\section{Trapdoored Matrices from the Learning Parity with Noise Assumption}\label{sec:lpn}

\subsection{The Learning Parity with Noise (LPN) Assumption}
\label{sec:lpnassm}

\def\Derr{\mathsf{Bin}}



We define a version of the learning parity with noise assumption~\cite{DBLP:conf/crypto/BlumFKL93,DBLP:conf/focs/Alekhnovich03,DBLP:conf/tcc/IshaiPS09} generalized over arbitrary finite fields. 
Let $\bbF$ be a finite field and let $\Derr_p$ be an error distribution over $\bbF$ that outputs a random non-zero element of $\bbF$ with probability $p$ and $0$ with probability $1-p$. Define the following distributions: 
\def\dlpn{\mathcal{D}_{\mathsf{lpn}}}
\def\dunif{\mathcal{D}_{\mathsf{unif}}}
\begin{align*}
 & \dlpn := \{ (\veca_i, b_i): \vecs \gets \bbF^k, \veca_i \gets \bbF^k, e_i \gets \Derr_p, b_i = \langle \veca_i, \vecs \rangle + e_i \}_{i\in [n]} \\
 &  \dunif := \{ (\veca_i, b_i): \veca_i \gets \bbF^k, b_i \gets \bbF \}_{i\in [n]}
\end{align*}
The learning parity with noise (LPN) assumption says that the two distributions above are computationally indistinguishable for any $n=\mathsf{poly}(k)$ samples.

\begin{conjecture}[LPN]\label{conj:lpn}
    The learning parity with noise (LPN) assumption with parameters $k,n=\mathsf{poly}(k),p$, conjectures that for all probabilistic $\mathsf{poly}(k)$-time adversaries $A$, there is a negligible function $\mu:\mathbb{N} \to \mathbb{N}$ such that:
$$ 
\Pr[A(\dlpn) = 1] - \Pr[A(\dunif) = 1] \leq \mu(k)
$$
\end{conjecture}

Note that the security (e.g. the runtime and the distinguishing advantage of $A$) is parameterized by the LPN dimension $k$. 
The LPN assumption (and its generalization over larger fields) has found a number of applications in cryptography, and seems exponentially hard in a wide range of parameters (see e.g. ~\cite{DBLP:conf/eurocrypt/LiuWYY24} and the references therein). 
The best algorithms known for LPN in dimension $k$ with $2^{O(k/\log k)}$ samples and $p=O(1)$ run in time $2^{O(k/\log k)}$~\cite{DBLP:journals/jacm/BlumKW03}; and with 
$\mathsf{poly}(k)$ samples runs in time $2^{O(k/\log \log k)}$~\cite{DBLP:conf/approx/Lyubashevsky05}. On the other hand, with $p=O((\log k)/k)$, the assumption is trivially broken in $\mathsf{poly}(k)$ time. In between, for a general $p$, the best known attacks run in time $2^{\tilde{O}(H(p)\cdot k)}$.

Conjecture~\ref{conj:lpn} is on the safer side for $p\gg (\log k)/ k$; the more aggressive subexponential LPN assumption conjectures the following, and is plausible for a polynomial $p = k^{-\beta}$ for any constant $\beta<1$ (with a related $\alpha$).

\begin{conjecture}[Subexponential LPN]\label{conj:subexp-lpn}
    The subexponential learning parity with noise (Subexp-LPN) assumption with parameters $k,n,p$, conjectures that there exists a constant $\alpha<1$ such that for $n \leq 2^{k^\alpha}$, for all probabilistic $2^{k^{\alpha}}$-time adversaries $A$:
$$ 
\Pr[A(\dlpn) = 1] - \Pr[A(\dunif) = 1] \leq 2^{-k^\alpha}
$$
\end{conjecture}

\subsection{Construction from LPN}
\begin{theorem}
  \label{thm:LPN}
   Under the polynomial hardness of the learning parity with noise assumption over a field $\mathbb{F}$, there is a collection of trapdoored matrices for which, given the trapdoor, $n\times n$ matrix-vector multiplication can be performed in time $n^{1+\epsilon}$ for an arbitrarily small constant $\epsilon>0$. Assuming sub-exponential hardness, this can be done in time 
  $n\cdot \mathsf{poly}(\log n)$.
\end{theorem}

\paragraph{The Base Construction.}
Fix an integer $k$ and a noise parameter $p \in [0,1]$ that we will choose subsequently.
A random matrix $\matM \in \bbF^{n\times m}$ from our distribution $\mathcal{D}_{n,k,p}$ is generated by sampling uniformly random $\matA \gets \bbF^{n\times k}$, $\matB \gets \bbF^{k\times n}$, and a Bernoulli matrix $\matE$ with entries drawn from $\mathsf{Bin}(p)$, and letting $$\matM = \matA \matB + \matE~.$$

The complexity of computing $\matM \vecv$ for an arbitrary $\vecv \in \bbF^n$ is
\begin{equation} \label{eqn:recursion}
T(n) = \frac{2n}{k} \cdot T(k) + O(n+pn^2) 
\end{equation}
since one first cuts $\vecv$ into $n/k$ pieces of length $k$ each, computes $n/k$ matrix-vector multiplications in $k$ dimensions followed by $n/k$ additions of $k$-dimensional vectors to compute $\matB \vecv$, and another $n/k$ matrix-vector multiplications in $k$ dimensions to compute $\matA \matB \vecv$. Computing $\matE \vecv$ takes another $pn^2$ operations, and adding the two takes $n$ operations.

The distinguishing advantage between $\matM \gets \mathcal{D}_{n,k,p}$ and a uniformly random matrix in $\bbF^{n\times n}$ is
$$ \delta(n) = n\cdot f(k,n,p)$$
where $f(k,n,p)$ is the advantage by which one can break LPN in $k$ dimensions with error parameter $p$ and $n$ samples.

Letting $T(k) = k^2$ and $p(k,n) = (\log k)^c/ k$  for a constant $c>1$ gives us
$$ T(n) = \frac{2n}{k} \cdot k^2 + O\bigg(n + \frac{n^2 (\log k)^c}{k} \bigg) = O\bigg(nk+ \frac{n^2 (\log k)^c}{k} \bigg)$$
Setting $k = \tilde{O}(\sqrt{n})$ gives us
$$T(n) = \tilde{O}(n^{3/2})~.$$
Pseudorandomness of $\matM$ relies on the plausible hardness of LPN in $k$ dimensions with $\tilde{O}(k^2)$ samples and noise rate $p = \frac{(\log k)^c}{k}$, which is conjectured to hold against $\mathsf{poly}(k)$-time adversaries as long as $c > 1$.

\paragraph{The Recursion.} Can we do better? We will run the recursion suggested by equation~(\ref{eqn:recursion}) by picking matrices $\matA$ and $\matB$ to be (trapdoored) pseudorandom matrices as well. That is, we will let the distribution $$\mathcal{D}_{i} := \mathcal{D}^{(i)}_{n_i,n_{i+1},p_{i}}$$ be a distribution over matrices $\matM_i \in \bbF^{n_i\times n_i}$ generated by 
\begin{itemize}
    \item  picking $n_{i+1} \times n_{i+1}$ matrices $\matA_{i,j} \gets \mathcal{D}_{i+1}$, for $j \in [n_{i}/n_{i+1}]$, and letting $$\matA_i := [\matA_{i,1} \sslash \matA_{i,2} \sslash \ldots \sslash \matA_{i,n_{i}/n_{i+1}}] \in \bbF^{n_i\times n_{i+1}}~;$$
    where $[\matA \sslash \matB]$ refers to concatenating the matrices $\matA$ and $\matB$ vertically.
    \item picking $n_{i+1} \times n_{i+1}$  matrices $\matB_{i,j} \gets \mathcal{D}_{i+1}$, for $j \in [n_{i}/n_{i+1}]$, and letting $$\matB_i := [\matB_{i,1} \| \matB_{i,2} \| \ldots \| \matB_{i,n_{i}/n_{i+1}}] \in \bbF^{n_{i+1}\times n_{i}}~;$$
    where where $[\matA \| \matB]$ refers to concatenating the matrices $\matA$ and $\matB$ horizontally.
    \item letting $\matM_i = \matA_i \matB_i + \matE_i$ where $\matE_i \gets (\mathsf{Bin}(p_i))^{n_i\times n_i}$.
\end{itemize}
We will choose $n_0 = n$, $n_{i+1} = n_i^{1-\epsilon}$ and $p_i = n_{i+1}^{-\delta}$ for constants $\epsilon > 0$ and $0 < \delta < 1$, generalizing the construction above in the natural way. The recursion equation~\ref{eqn:recursion} then tells us that
\begin{align*} 
T(n_i) & = \frac{2n_{i}}{n_{i+1}}  \cdot T(n_{i+1})+ O(n_{i} + p_i n_i^2) \\
& = 2n_i^{\epsilon} \cdot T(n_{i}^{1-\epsilon}) + O\big(n_i^{\mathsf{max}(1,2-\delta(1-\epsilon))}\big)
\end{align*}
Running the recursion for $\ell$ steps gives us matrices of dimension $n^{(1-\epsilon)^\ell}$ whereby we apply the trivial matrix-vector multiplication algorithm of complexity $n^{2\cdot (1-\epsilon)^\ell}$. To compute the runtime, observe that the first term in the recursion, after $\ell$ steps becomes 
$$ n^\epsilon \cdot  n^{\epsilon(1-\epsilon)} \cdot n^{\epsilon(1-\epsilon)^2} \cdot \ldots \cdot n^{\epsilon(1-\epsilon)^{\ell-1}} \cdot n^{2\cdot (1-\epsilon)^\ell}  \leq n^{\epsilon \cdot \frac{1}{\epsilon} + 2\cdot (1-\epsilon)^\ell} = n^{1+2\cdot (1-\epsilon)^\ell} $$
where the first inequality is by upper-bounding the geometric sum $\sum_{j=0}^\ell (1-\epsilon)^j$ by $1/\epsilon$. Under the subexponential LPN assumption, we can only go as far as setting $n_\ell = n^{(1-\epsilon)^\ell}$ to be $\mathsf{poly}(\log n)$, which means this sum will be $n\cdot \mathsf{poly}(\log n)$. The second term of the sum becomes 
$$ n^c + n^{\epsilon+(1-\epsilon)\cdot c} + n^{\epsilon + \epsilon\cdot (1-\epsilon) + (1-\epsilon)^2\cdot c} + \ldots + n^{\epsilon\cdot \sum_{j=0}^{\ell-1} (1-\epsilon)^j + (1-\epsilon)^{\ell} c} $$
where $c := \mathsf{max}(1,2-\delta\cdot (1-\epsilon)) > 1$.
Note the exponents decrease monotonically since $c>1$, and so the sum is dominated by the first term. Setting $\delta \approx 1$ and $\epsilon \approx 0$ gives us $c \approx 1$, and so the sum is $O(n\cdot \mathsf{poly}(\log n))$.

Since the base case is LPN in $\mathsf{poly}(\log n)$ dimensions with an inverse polynomial error rate, under the subexponential assumption, the distinguishing advantage is negligible in $n$. 

If we only assume the polynomial LPN assumption, on the other hand, we will stop the recursion at $n_\ell = n^{\gamma}$ for an arbitrarily small constant $\gamma > 0$. Thus will result in a complexity $T(n)$ that is $n^{1+O(\gamma)}$ for an arbitraily small positive constant $\gamma$.

\begin{corollary}
    \label{cor:matmult}
     Under the sub-exponential hardness of the learning parity with noise assumption over a field $\mathbb{F}$, there is a collection of trapdoored matrices $\matM$ for which, given the trapdoor, $n\times n$ matrix-matrix multiplication with any matrix $\matV$ can be performed in time 
  $n^2\cdot \mathsf{poly}(\log n)$. 
\end{corollary}

\paragraph{Left and Right Multiplication.}
It is easy to see that our pseudorandom matrix $\matM = \matA \matB + \matE$ supports both left and right multiplication in near linear time as long as the constituent submatrices of $\matA$ and $\matB$ have the same property recursively. It is unclear, though, whether one can modify the construction to support quick multplication of a vector by $\matM^{-1}$ as well as $\matM$.


\section{Trapdoored Matrices from the McEliece Assumption}\label{sec:mcel}

\subsection{The McEliece Assumption Family} 
\label{sec:mceassm}
Let $\matG \in \bbF^{n\times k}$ be the (fixed, publicly known) generator matrix of an $[n,k,d]$ linear code $\mathcal{C}$. The McEliece assumption~\cite{McEliece1978} w.r.t. $\mathcal{C}$ conjectures that the following two distributions on matrices are computationally indistinguishable:
\def\dmce{\mathcal{D}_{\mathsf{McEliece}}}
\begin{align*}
 & \dmce := \{ \matM = \matP \matG \matS: \matS \gets \bbF^{k\times k}, \matP \gets \mathsf{Perm}(n)\} \\
 &  \dunif := \{ \matM \gets \bbF^{n\times k}\}
\end{align*}
where $\mathsf{Perm}(n)$ is the set of all $n\times n$ permutation matrices.\footnote{
For a reader familiar with the McEliece encryption scheme, we remark that the assumption as formulated above is {\em sufficent} for the security of the McEliece cryptosystem (in conjunction with the LPN assumption over $\bbF$), but is not necessary for it.}

McEliece formulated the assumption with $\mathcal{C}$ being the Goppa code~\cite{McEliece1978}; this still remains unbroken and is the basis of candidate encryption schemes submitted to the NIST post-quantum cryptography standardization process~\cite{ClassicMcE}. The McEliece assumption w.r.t. ``quasi-cyclic'' codes has also been considered in the literature~\cite{DBLP:conf/isit/MisoczkiTSB13}, and is the basis of a different NIST encryption candidate~\cite{bike}. All these choices of codes permit {\em near-linear time encoding}: that is, given a vector $\vecv \in \bbF^k$, one can compute $\matG \vecv \in \bbF^n$ in time $\tilde{O}(n+k)$, as opposed to the trivial $O(nk)$. On the other hand, when $\mathcal{C}$ is either the Reed-Solomon or the Reed-Muller code, the assumption is broken via an attack of Sidelnikov and Shestakov~\cite{SS92}. In the case of Goppa codes, despite recent attempts at cryptanalysis (see, e.g. \cite{DBLP:journals/tit/FaugereGOPT13,DBLP:conf/eurocrypt/CouvreurOT14,DBLP:conf/asiacrypt/CouvreurMT23,DBLP:journals/tit/BardetMT24}), the best known attack from earlier this year incurs very slightly subexponential complexity~\cite[Theorem~3]{DBLP:journals/corr/abs-2407-15740}.

Our assumption will be that the McEliece assumption holds w.r.t. a code family that permits near-linear-time encoding, e.g. the Goppa or quasi-cyclic codes~\cite{ClassicMcE,bike,DBLP:conf/isit/MisoczkiTSB13}. As in the case of the LPN assumption, we can assume either the (conservative) polynomial or the (aggressive) subexponential versions of the assumption, with different results.


\subsection{Construction from McEliece}

\begin{theorem}
  \label{thm:McEliece}
  Under the polynomial hardness of the McEliece assumption over a field $\mathbb{F}$, there is a collection of trapdoored matrices for which, given the trapdoor, $n\times n$ matrix-vector multiplication can be performed in time $n^{1+\epsilon}$ for an arbitrarily small constant $\epsilon>0$. Assuming sub-exponential hardness, this can be done in time $n\cdot \mathsf{poly}(\log n)$. 
\end{theorem}

\paragraph{The Construction.} We will describe the base construction that achieves a runtime of $\tilde{O}(n^{3/2})$ here. A recursion of the exact same form as in the LPN setting gives us either $n^{1+\epsilon}$ or $n\cdot \mathsf{poly}(\log n)$, under the polynomial, resp. subexponential McEliece assumption.

Let $\matG$ be the $n\times k$ generator matrix of any near-linear-time encodable code $\mathcal{C}$. That is, given a vector $\vecv \in \bbF^k$, one can compute $\matG \vecv \in \bbF^n$ in time $\tilde{O}(n+k)$. Our trapdoored matrix will be the matrix $\matM = \matP \matG \matS$ where $\matP$ is a random $n\times n$ permutation matrix and $\matS$ is a random $k\times k$ matrix, that comes from the McEliece assumption.

Computing $\matM \vecv$ proceeds by iterative multiplication: first with $\matS$ taking time $O(k^2)$; then with $\matG$ taking time $\tilde{O}(k+n)$ using the linear-time encodability of the code; and finally with $\matP$ taking time $O(n)$. The total runtime is $\tilde{O}(k^2 + n)$, in contrast to the trivial $O(kn)$ time multiplication. 

To construct a square $n\times n$ trapdoored matrix, we stack $n/k$ instances of this construction side by side resulting in a matrix $\matM = [\matM_1 || \ldots || \matM_{n/k}]$ where each $\matM_i \in \bbF^{n\times k}$ is constructed as above. Now, the total runtime of a matrix-vector multiplication is 
$$ O\bigg( \frac{n}{k} \cdot (k^2+n) + \frac{n}{k} 
 \cdot n\bigg)  = O\bigg(nk + \frac{n^2}{k} \bigg)$$
 where the first term is the time to perform the above algorithm $n/k$ times, and the second is the time to aggregate the answers. Setting $k = \sqrt{n}$ already gives us subquadratic time. A recursion on constructing $\matS$, similar to the LPN case, gives us quasilinear time. 



\section{Trapdoored Matrices over the Reals}\label{sec:reals}

Many of the applications of random matrices use Gaussian or sub-Gaussian matrices over the reals, rather than matrices over finite fields. 
Unfortunately, cryptographic hardness assumptions concerning non-periodic real numbers are scarce.
We thus introduce a computational hardness assumption regarding a well-studied mathematical object. We motivate this conjecture in Section~\ref{sec:reals}.

\subsection{Construction from Kac's Random Walks}
\label{sec:kacassm}

We introduce a conjecture on Kac's random walks, which immediately yields a simple construction of trapdoored matrices over the reals for the Haar-measure (i.e., the uniform distribution over rotation matrices).
We then use this family of trapdoored matrices to construct another family for \emph{any} Haar-invariant distribution. 

In his 1954 physics paper, Kac~\cite{kac1956foundations} introduced a random walk on the sphere as a model for a Boltzmann gas (for a summary on the influence of this model on kinetic theory, see~\cite{{mischler2013kac}}). 
This is a Markov chain in which the time is discrete, but the state space is continuous. 
We present it in terms of linear transformations.
Let~$Q_0 = I_n = \mathbb{R}^{n \times n}$ be the~$n$-dimensional identity matrix.
For any pair of distinct indices~$i,j\in [n]$ and any angle~$\theta\in [0,2\pi)$ we denote by~$R_{i,j,\theta}$ the rotation of the plane spanned by the basis vectors~$e_i,e_j$ by angle~$\theta$, that is, the sole linear transformation for which
\begin{align*}
    &R_{i,j,\theta}(e_k) = e_k  &\forall k\notin\{i,j\}\\
    &R_{i,j,\theta}(e_i) = \cos \theta \cdot e_i + \sin \theta \cdot e_j \; &\\
    &R_{i,j,\theta}(e_j) = -\sin \theta \cdot e_i + \cos \theta \cdot e_j \;. &
\end{align*}
We define a random walk starting from~$Q_0=I_n$ in which every step consists of a multiplication with a matrix~$R_{i,j,\theta}$ where the indices~$i,j$ and the angle~$\theta$ are drawn uniformly. That is~$Q_{t} \leftarrow R_{i_t,j_t,\theta_t}$. The resulting matrix is the state~$Q_T$ of the random walk after~$T$ steps.
Essentially, in each step, we perform a random rotation in low-dimension. It is thus reasonable to expect that after enough steps we will converge to a random ``rotation'' of the entire space.

Understanding the mixing time of the Kac's random walk over the sphere (that is, considering the evolution of only a single column of~$Q_t$, or equivalently of~$Q_t \cdot e_1$) already required significant mathematical effort. In a sequence of breakthrough papers~\cite{janvresse2001spectral,mischler2013kac,pillai2017kac} it was proven that when~$T=\Theta(n \log n)$ the distribution of the resulting state~$Q_t \cdot e_1$ approaches the uniform distribution of~$S^{n-1}$, even in the total variation distance.
This is the best possible mixing time due to a simple lower coupon collector-style lower bound. 

Understanding the mixing time of the entire transformation~$Q_t$ is even more challenging. Although it is easy to show that the chain converges to the uniform distribution in the special orthogonal group~$SO(n)$, only in 2012, Jiang~\cite{jiang2012total} showed that the chain mixes in polynomial time. The bounds have since been improved and it is currently known that the mixing time $T(n)$ is between~$\Omega(n^2) \leq T(n) \leq O(n^4 \log n)$.

We pose the following natural conjecture, stating that after~$n\cdot \poly(\log n)$ steps, the resulting state~$Q_T$ should already be \emph{computationally indistinguishable} from the uniform distribution over~$SO(n)$. As stated above, after that number of steps any small set of columns is distributed as i.i.d. samples from~$S^{n-1}$, yet the distribution of the entire matrix is still not \emph{statistically} close to uniform. 
We are not aware of previous conjectures or works considering the computational pseudo-randomness of these walks.

\begin{conjecture}\label{conj:kac}
    There exists some~$T=n \cdot \poly\log(n)$ such that the distribution of~$Q_T$ is computationally indistinguishable from the Haar-measure over~$SO\left(n\right)$.
\end{conjecture}

We note that many existing hardness assumptions can also be phrased as running a mixing Markov chain for fewer steps than necessary for statistical convergence and conjecturing that the resulting state is pseudo-random.
Further supporting this conjecture is a recent result~\cite{jain2022fast} in which it was shown that the transformation~$Q_T$ with~$T=\Theta(n \log n)$ suffices for recovering the JL Lemma. This is an intricate unconditional proof that~$Q_T$ preserves all pair-wise distances of a point-set w.h.p., recovering the previous results on Fast JL Transforms~\cite{ailon2013almost}.
We conjecture that the same distribution is in fact \emph{pseudorandom}, and would thus suffice for the far more general applications as well.

Assuming this conjecture, a construction of trapdoored matrices for the uniform distribution over~$SO(n)$ is straightforward: Every step of the Kac's walk affects only two coordinates and can thus be applied in~$O(1)$ time to a vector. 
Furthermore, as each step is easily invertible we are also able to quickly apply the inverse matrix~$Q_T^{-1}$.

\subsection{Trapdoored Matrices over any Haar-Invariant Distribution}
We next present a reduction showing that a trapdoored matrix family for the Haar-measure suffices to construct one for any distribution which is either Haar-invariant or symmetrically Haar-invariant.

\begin{theorem}\label{thm:haarinv}
Let~$\mathcal{D}$ be \emph{any} distribution that is either Haar-invariant or symmetrically Haar-invariant.
  Under the psuedorandomness of Kac's random walks, there is a collection of trapdoored matrices over the distribution~$\mathcal{D}$ for which, given the trapdoor, $n\times n$ matrix-vector multiplication can be performed in time $n\cdot \mathsf{poly}(\log n)$. 
\end{theorem}

This generalization follows from the following observation, which informally states that rotationally-invariant matrix distributions are fully characterized by their singular values or eigenvalues.

\begin{lemma}\label{lem:haar}
    Let~$\mathcal{D}$ be a Haar-invariant distribution.
    Then, there exists a distribution~$\Bar{\mathcal{D}}$ of \textbf{diagonal} matrices such that if we sample~$O_1,O_2\sim SO(n)$ and~$D\sim \Bar{\mathcal{D}}$ independently, then~$O_1\cdot D\cdot O_2$ is distributed as~$\mathcal{D}$.
    Furthermore, if we can sample from~$\mathcal{D}$ in~$T(n)$ time then we can sample from~$\Bar{\mathcal{D}}$ in~$O\left(T\left(n\right)+n^3\right)$ time.
\end{lemma}
\begin{proof}
    Let~$A\sim \mathcal{D}$ be a random variable. 
    There always exists a Singular Value Decomposition (SVD) of~$A$ into a product of the form~$A=U\cdot D\cdot V^t$ such that~$D$ is a diagonal matrix and~$U,V\in SO(n)$.
    Let~$O_1,O_2\sim SO(n)$ be two Haar-distributed matrices sampled independently. 
    Due to~$\mathcal{D}$ being Haar-invariant,~$O_1 \cdot A \cdot O_2$ is distributed as~$\mathcal{D}$.
    On the other hand, we observe that~\[
    O_1 \cdot A \cdot O_2 = O_1 \cdot U\cdot D\cdot V^t \cdot O_2 = 
    (O_1 \cdot U)\cdot D\cdot (V^t \cdot O_2),
    \]
    where $(O_1 \cdot U), (V^t \cdot O_2)$ are distributed as two independent samples from the Haar-measure.
    In particular, if we denote the distribution of~$D$ by~$\Bar{\mathcal{D}}$ then the Lemma's statement follows.
    Furthermore, we can sample~$D\sim \Bar{\mathcal{D}}$ by sampling~$A\sim \mathcal{D}$ and computing the diagonal part~$D$ resulting from the SVD of~$A$.
\end{proof}

Lemma~\ref{lem:haar} implies Theorem~\ref{thm:haarinv} for Haar-invariant distributions~$\mathcal{D}$:
To sample a matrix and a trapdoor from our new trapdoored family, we would sample a diagonal matrix~$D\sim \Bar{\mathcal{D}}$ as well as two rotation matrices~$O_1,O_2$ independently from the trapdoored matrix family over the Haar-measure constructed in the beginning of the section.
Our sampled matrix is~$A:=O_1\cdot D\cdot O_2$ which due to Lemma~\ref{lem:haar} and the definition of trapdoored families, is distributed indistinguishably from~$\mathcal{D}$.
To quickly multiply the matrix~$A$ by a vector~$v$, we break the multiplication into three steps: first, we multiply~$O_2 \cdot v$, which we can do quickly due to~$O_2$ being from a trapdoored family; second, we multiply~$D\cdot\left(O_2 v\right)$, which we can do in linear time as~$D$ is a diagonal matrix; finally, we multiply~$O_1\cdot \left(DO_2 v\right)$ using~$O_1$'s trapdoor.

We next provide a similar construction for symmetrically Haar-invariant distributions.
\begin{lemma}\label{lem:symhaar}
    Let~$\mathcal{D}$ be a symmetrically Haar-invariant distribution.
    Then, there exists a distribution~$\Bar{\mathcal{D}}$ of \textbf{diagonal} matrices such that if we sample~$O\sim SO(n)$ and~$D\sim \Bar{\mathcal{D}}$ independently, then~$O\cdot D\cdot O^{-1}$ is distributed as~$\mathcal{D}$.
    Furthermore, if we can sample from~$\mathcal{D}$ in~$T(n)$ time then we can sample from~$\Bar{\mathcal{D}}$ in~$O\left(T\left(n\right)+n^3\right)$ time.
\end{lemma}
\begin{proof}
    Let~$A\sim \mathcal{D}$ be a random variable. 
    Since~$A$ is symmetric, there always exists an eigendecomposition of~$A$ into a product~$A=U\cdot D\cdot U^{-1}$ such that~$D$ is a diagonal matrix and~$U\in SO(n)$.
    Let~$O\sim SO(n)$ be a Haar-distributed matrix. 
    Due to~$\mathcal{D}$ being Haar-invariant,~$O \cdot A \cdot O^{-1}$ is distributed as~$\mathcal{D}$.
    On the other hand, we observe that~\[
    O \cdot A \cdot O^{-1} =
    O \cdot U\cdot D\cdot U^{-1} \cdot O^{-1} = 
    (O\cdot U) \cdot D \cdot (O\cdot U)^{-1}
    ,
    \]
    where $(O \cdot U)$ is distributed according to the Haar-measure.
    In particular, if we denote the distribution of~$D$ by~$\Bar{\mathcal{D}}$ then the Lemma's statement follows.
    Furthermore, we can sample~$D\sim \Bar{\mathcal{D}}$ by sampling~$A\sim \mathcal{D}$ and computing the diagonal part~$D$ resulting from the eigendecomposition of~$A$.
\end{proof}

Lemma~\ref{lem:symhaar} implies Theorem~\ref{thm:haarinv} for symmetrically Haar-invariant similarly to the non-symmetric case, with one difference: We observe that this time we need a quick matrix-vector multiplication not only with~$O$ but also with~$O^{-1}$, which we do get from the construction in the beginning of the section.

\paragraph{Remark:} 
In the general reduction above, the matrix-vector multiplication time is near-linear yet the sample time from the trapdoored matrix distribution is~$T(n)+O(n^3)$ where~$T(n)$ is the sample time from~$\mathcal{D}$.
We note that for most of the ``useful" distributions~$\mathcal{D}$ (such as that of i.i.d. normal coefficients or GOE), directly sampling from~$\Bar{\mathcal{D}}$ is actually much quicker as the eigenvalues (or singular values) of these distributions are well-understood and can be directly sampled. 

\section{Worst Case to Average Case Reductions for Linear Algebra}\label{sec:lingalg}

\subsection{Uniform Error Correction for Matrix Multiplication}
We begin by describing a simple worst-case to average-case reduction for matrix multiplication, for which \emph{any} positive success probability in the average-case suffices. 
All of the following reductions can be instantiated with the trapdoored matrix families constructed in Sections~\ref{sec:lpn} or~$\ref{sec:mcel}$.

\begin{theorem}
    Let~$\mathbb{F}_p$ be a finite field of order~$p$ and let~$\varepsilon>0$. Assume the existence of a trapdoored matrix family over~$\mathbb{F}_p$.
    Let~$\mathcal{M} : \mathbb{F}_p^{n\times n} \times \mathbb{F}_p^{n\times n} \rightarrow \mathbb{F}_p^{n\times n}$ be an algorithm running in time~$T(n)$ such that for every sufficiently large~$n$, 
    \[
    \Pr_{A,B\sim \mathbb{F}_p^{n\times n}} \Big[
    A\cdot B = \mathcal{M}\left(A,B\right) 
    \Big] \geq\varepsilon.
    \]
    Then, we can construct an algorithm that for \textbf{any} matrices $A,B\in \mathbb{F}_p^{n\times n}$ computes the multiplication~$A\cdot B$ with high probability, in time~$\tilde{O}\left(T(n)\right)$.
\end{theorem}
\begin{proof}
    Let~$A,B$ be two input matrices.
    Let~$R,Q^t$ be matrices drawn from our trapdoored family over~$\mathbb{F}_p^{n\times n}$.
    We observe that~$A+R, B+Q$ are (indistinguishable from) uniformly distributed matrices, and that $$AB = (A+R)(B+Q)-AQ-RB-RQ~.$$
    Furthermore, each of the multiplications~$AQ,RB,RQ$ can be computed in~$\tilde{O}(n^2)$ time using the trapdoor of either~$R$ or~$Q^t$.
    In particular, we can compute $$\mathcal{M}(A+R,B+Q)-AQ-RB-RQ$$ 
    in~$T(n)+\tilde{O}(n^2)$ time, which exactly equals~$AB$ with probability at least~$\varepsilon$.
    We can verify whether the resulting matrix equals~$AB$ or not with high probability in~$\tilde{O}(n^2)$ time using Freivalds' algorithm.
    By repeating the above~$\Theta(\log n)$ times we end up with~$AB$ with high probability.
\end{proof}

With a slight modification of this simple reduction, we also obtain error-correction of any matrix multiplication algorithm that on average agrees with slightly more entries of the resulting matrix than a random matrix would. 

\begin{theorem}\label{thm:errcorrectMM}
    Let~$\mathbb{F}_p$ be a finite field of order~$p$ and let~$\varepsilon>0$. Assume the existence of a trapdoored matrix family over~$\mathbb{F}_p$.
    Let~$\mathcal{M} : \mathbb{F}_p^{n\times n} \times \mathbb{F}_p^{n\times n} \rightarrow \mathbb{F}_p^{n\times n}$ be an algorithm running in time~$T(n)$ such that for every sufficiently large~$n$, 
    \[
    \mathbb{E}_{A,B\sim \mathbb{F}_p^{n\times n}} \Big[
    \text{dist}\left(A\cdot B, \ \mathcal{M}\left(A,B\right)\right) 
    \Big] \leq\left( 1-\frac{1}{p}-\varepsilon\right)n^2.
    \]
    Then, we can construct an algorithm that for \textbf{any} matrices $A,B\in \mathbb{F}_p^{n\times n}$ computes the multiplication~$A\cdot B$ \textbf{exactly} with high probability, in time~$\tilde{O}\left(T(n)\right)$.
\end{theorem}

\begin{proof}
    Denote by~$\mathcal{E}:=\mathcal{M}(R,Q)-RQ$, for uniformly drawn~$R,Q$, the matrix of \emph{errors} produced by~$\mathcal{M}$. We know that the expected Hamming weight of~$\mathcal{E}$ is at most~$\left(1-\frac{1}{p}-\varepsilon\right)n^2$.
    Let~$P,P'$ be two uniformly drawn permutation matrices, and let~$D$ be a diagonal matrix in which each diagonal element is uniformly drawn from~$\mathbb{F}_p\setminus\{0\}$.
    For any pair of indices~$(i,j)\in [n]^2$, we have that 
    \[
    \left(DP\mathcal{E}P'\right)_{i,j} = 
    \begin{cases}
        0, \ \text{with probability $q$} \\
        k, \ \text{with probability }\frac{1-q}{p-1}\text{ for }k\in \mathbb{F}_p\setminus\{0\},
    \end{cases}
    \]
    for some~$q\geq \frac{1}{p}+\varepsilon$.

    We can thus draw~$R,Q^t$ from our trapdoored family, and the permutation matrices~$P,P'$ and the diagonal one~$D$ uniformly.
    Then, we can compute \[
    D\cdot P \cdot \mathcal{M}\left(P^{-1} D^{-1}\left(A+R\right),\left(B+Q\right)\left(P'\right)^{-1}\right)\cdot P'-AQ-RB-RQ
    \;\; = \;\;
    AB + DP\mathcal{E}P'
    \]
    in~$T(n)+\tilde{O}(n^2)$ time, and observe that each entry~$(i,j)$ of the resulting matrix equals~$(AB)_{i,j}$ with probability~$q$ and is otherwise uniformly distributed among the other field elements.
    In particular, by repeating the above process~$\Omega\left(\frac{\log \left(pn\right)}{\varepsilon^2 p}\right)$ times and taking the plurality in each matrix entry we end up with~$AB$ with high probability, by a union bound over the entries.
\end{proof}

\subsection{Other Linear Algebraic Operations}
We describe the first reductions from worst-case to average-case for linear algebraic operations in addition to matrix multiplication.

Let~$\mathbb{F}_p$ be a finite field of order~$p$.
We denote by~$\mathbb{F}_p^{n\times n}$ and by~$GL_n(\mathbb{F}_p)$ the set of all~$n$-by-$n$ matrices over~$\mathbb{F}_p$ and the set of all non-singular such matrices, respectively. We abuse notation by using the same notation for the uniform distribution over each set.
We denote by~$$q_p := 1 - \inf_{n\in \mathbb{N}} \frac{|GL_n(\mathbb{F}_p)|}{|\mathbb{F}_p^{n\times n}|}$$ the lowest upper bound on the probability that a random matrix over~$\mathbb{F}_p$ is singular.
It is known that~$q_2 \approx 0.711, q_3\approx 0.440, q_4\approx 0.311$ and that generally~$q_p < \frac{1}{p-1}$.

\paragraph{Matrix Inversion:}
Given an algorithm that on average over~$GL_n(\mathbb{F}_p)$ successfully inverts a matrix with any positive probability, we construct an algorithm that inverts a worst-case matrix with high probability.
\begin{theorem}
    Let~$\mathbb{F}_p$ be a finite field of order~$p$ and let~$\varepsilon>0$. Assume the existence of a trapdoored matrix family over~$\mathbb{F}_p$ supporting two-sided multiplications.
    Let~$\mathcal{M} : \mathbb{F}_p^{n\times n} \rightarrow \mathbb{F}_p^{n\times n}$ be an algorithm running in time~$T(n)$ such that for every sufficiently large~$n$, 
    \[
    \Pr_{A\sim GL_n(\mathbb{F}_p)} \Big[
    A^{-1} = \mathcal{M}\left(A\right) 
    \Big] \geq\varepsilon.
    \]
    Then, we can construct an algorithm that for \textbf{any} matrix $A\in GL_n(\mathbb{F}_p)$ computes the inverse~$A^{-1}$ with high probability, in time~$\tilde{O}\left(T(n)\right)$.
\end{theorem}
\begin{proof}
    Let~$R\sim \mathbb{F}_p^{n\times n}$ be a uniformly drawn random matrix, and let~$D$ be the event that~$R\in GL_n(\mathbb{F}_p)$. Conditioned on~$D$, the matrix~$R$ is uniformly distributed over~$GL_n(\mathbb{F}_p)$.
    We draw~$R$ from our trapdoored matrix family and compute~$R\cdot \mathcal{M}(A\cdot R)$. Both matrix multiplications are performed using the trapdoor, which allows multiplication by~$R$ from either side.
    Conditioned on~$D$, the distribution of~$A\cdot R$ is indistinguishable from uniform over~$GL_n(\mathbb{F}_p)$ and in particular with probability~$\geq \varepsilon$ the algorithm~$\mathcal{M}$ returns~$(AR)^{-1} = R^{-1}A^{-1}$ and in turn our computed matrix equals~$A^{-1}$. We can verify whether the computed matrix is indeed~$A^{-1}$ using Freivalds' algorithm (used to verify whether the result~$B$ satisfies~$A\cdot B = I$). As~$\Pr(D)=1-q_p$, repeating the above for~$\Theta\left(\frac{\log n}{(1-q_p)\varepsilon}\right)$ times suffices to succeed with high probability.
\end{proof}

We {remark} that we could alternatively define the success probability of~$\mathcal{M}$ over a uniform matrix from~$\mathbb{F}_p^{n\times n}$ rather than only invertible ones, defining a correct answer as either~$A^{-1}$ or a declaration that~$A$ is singular. If the success probability of~$\mathcal{M}$ in this setting is at least~$q_p+\varepsilon$ then its restriction to~$GL_n(\mathbb{F}_p)$ satisfies the conditions of the current theorem.
On the other hand, this is the best possible as the uninspiring algorithm that simply always answers ``singular" succeeds with probability~$q_p\pm o(1)$.

\paragraph{Solving Linear Equations:} Given an algorithm that correctly solves a random system of linear equations with a unique solution, with any positive probability, we construct an algorithm that works on any such system with high probability.
\begin{theorem}
    Let~$\mathbb{F}_p$ be a finite field of order~$p$ and let~$\varepsilon>0$. Assume the existence of a trapdoored matrix family over~$\mathbb{F}_p$ supporting two-sided multiplications.
    Let~$\mathcal{M} : \mathbb{F}_p^{n\times n} \times \mathbb{F}_p^n \rightarrow \mathbb{F}_p^{n}$ be an algorithm running in time~$T(n)$ such that for every sufficiently large~$n$, 
    \[
    \Pr_{A\sim GL_n(\mathbb{F}_p), \ b\sim \mathbb{F}_p^n} \Big[
    A\cdot \mathcal{M}\left(A,b\right) = b 
    \Big] \geq\varepsilon.
    \]
    Then, we can construct an algorithm that for \textbf{any} system $A\in GL_n(\mathbb{F}_p), b\in \mathbb{F}_p^n$ computes the unique solution~$x$ to the system~$Ax=b$ with high probability, in time~$\tilde{O}\left(T(n)\right)$.
\end{theorem}

\begin{proof}
    Let~$R,Q\sim \mathbb{F}_p^{n\times n}$ be uniformly drawn random matrices, and let~$D$ be the event that both~$R,Q\in GL_n(\mathbb{F}_p)$. Conditioned on~$D$, the matrices~$R,Q$ are uniformly distributed over~$GL_n(\mathbb{F}_p)$.
    We draw~$R,Q$ from our trapdoored matrix family and compute~$Q\cdot \mathcal{M}(R\cdot A\cdot Q, \ R\cdot b)$. All multiplications are performed using the trapdoor, which allows multiplication by~$R,Q$ from either side.
    Conditioned on~$D$, the distribution of~$Rb$ is indistinguishable from uniform over~$\mathbb{F}_p^n$, and that of~$RAQ$ is indistinguishable from uniform over~$GL_n(\mathbb{F}_p)$ and independent of that of~$Rb$ (due to~$Q$).
    In particular with probability~$\geq \varepsilon$ the algorithm~$\mathcal{M}$ returns~$(RAQ)^{-1} Rb = Q^{-1}A^{-1}b$ and in turn our computed solution vector equals~$A^{-1}b$. We can verify whether the solution is correct in~$O(n^2)$ time. As~$\Pr(B)=(1-q_p)^2$, repeating the above for~$\Theta\left(\frac{\log n}{(1-q_p)^2\varepsilon}\right)$ times suffices to succeed with high probability.
\end{proof}

\paragraph{Determinant Computation:}
Unlike all previous operations (matrix multiplication, matrix inversion, and solving a linear system), a solution to the determinant problem cannot be quickly verified. 
In turn, we only present a reduction from worst-case determinant computation to an average-case algorithm that succeeds with a sufficiently high probability. We later discuss the required success probability in more detail.

\begin{theorem}
    Let~$\mathbb{F}_p$ be a finite field of order~$p$. Assume the existence of a trapdoored matrix family over~$\mathbb{F}_p$.
    Let~$\mathcal{M} : \mathbb{F}_p^{n\times n} \rightarrow \mathbb{F}_p$ be an algorithm running in time~$T(n)$ such that for every sufficiently large~$n$, 
    \[
    \Pr_{A\sim \mathbb{F}_p^{n\times n}} \Big[
     \mathcal{M}\left(A\right) = det(A) 
    \Big] \geq 0.99.
    \]
    Then, we can construct an algorithm that for \textbf{any} matrix $A\in \mathbb{F}_p^{n\times n}$ computes $det(A)$ with high probability, in time~$\tilde{O}\left(T(n)\right)$.
\end{theorem}

We provide two different algorithms and proofs for fields of order~$p>2$ and for~$\mathbb{F}_2$.

\begin{lemma}
    Let~$\mathbb{F}_p$ be a finite field of order~$p>2$ and let~$\varepsilon>0$. Assume the existence of a trapdoored matrix family over~$\mathbb{F}_p$.
    Let~$\mathcal{M} : \mathbb{F}_p^{n\times n} \rightarrow \mathbb{F}_p$ be an algorithm running in time~$T(n)$ such that for every sufficiently large~$n$, 
    \[
    \Pr_{A\sim \mathbb{F}_p^{n\times n}} \Big[
     \mathcal{M}\left(A\right) = det(A) 
    \Big] \geq \frac{3+2q_p}{4}+\varepsilon .
    \]
    Then, we can construct an algorithm that for \textbf{any} matrix $A\in \mathbb{F}_p^{n\times n}$ computes $det(A)$ with high probability, in time~$\tilde{O}\left(T(n)\right)$.
\end{lemma}
\begin{proof}
    Initialize a list~$L$.
    For~$t=\Theta\left(\log \left(pn\right) /\varepsilon^2\right)$ times we repeat the following: Draw a matrix~$R$ from our trapdoored family, and query the algorithm on~$\mathcal{M}(R)$ and on~$\mathcal{M}(RA)$;
    If~$\mathcal{M}(R)=0$ then continue to the next iteration, otherwise append~$\mathcal{M}(RA)/\mathcal{M}(R)$ to the list~$L$.
    Finally, if at least a~$(1/2+\varepsilon)t$ entries in~$L$ are the same element then return this element, otherwise return~$0$.

    We prove the correctness in two parts, depending on the singularity of the input matrix~$A$.
    If~$A$ is invertible, then the probability that~$R$ is invertible and that~$\mathcal{M}$ answers correctly on both~$R$ and~$RA$ is by union bound at least 
    \[
    1-q_p - 2\left(1- \frac{3+2q_p}{4}-\varepsilon\right) = 
    \frac{1}{2}+2\varepsilon.
    \]
    Under these conditions,~$\mathcal{M}(RA)/\mathcal{M}(R)=det(RA)/det(R)=det(A)$ appears in more than~$(1/2+\varepsilon)t$ entries of~$L$ with high probability, and thus the returned value is correct.

    If~$A$ is singular, then we observe that for a random invertible matrix~$R$ the random matrix~$RA$ is independent of~$det(R)\sim \mathbb{F}_p\setminus\{0\}$.
    We can see this by picking a basis to~$\mathbb{F}_p^n$ in which the first~$k:=\dim \text{Ker}(A) \geq 1$ vectors span the null space of~$A$. A random invertible matrix~$R$ can be drawn with respect to this alternative basis (as both the distribution of such matrices and the determinant are invariant to the choice of basis). In this basis, the last~$n-k$ vectors fully determine~$RA$, yet even conditioned on all but the first basis vector the determinant of~$R$ is still uniform over~$\mathbb{F}_p\setminus\{0\}$.
    Therefore, despite~$RA$ not being distributed uniformly, we nonetheless have that conditioned on~$det(R)\neq0$, the value of~$\mathcal{M}(RA)/det(R)$ is either~$0$ (if~$\mathcal{M}(RA)=0$) or otherwise distributes uniformly over~$\mathbb{F}_p\setminus\{0\}$.
    In particular, if we know~$det(R)$ then the probability of adding each non-zero element to~$L$ is at most~$\frac{1-q_p}{p-1}$.
    As~$R$ \emph{is} distributed uniformly, then the probability of~$\det(R)\neq \mathcal{M}(R)$ is at most~$\frac{1}{4}-\frac{q_p}{2}-\varepsilon$ which is smaller than~$\frac{1}{2}-\frac{1-q_p}{p-1}$ for all~$p>2$. Hence, no non-zero value will appear~$(1/2+\varepsilon)t$ times in~$L$, with high probability.
\end{proof}

\begin{lemma}
    Let~$\varepsilon>0$. Assume the existence of a trapdoored matrix family over~$\mathbb{F}_2$.
    Let~$\mathcal{M} : \mathbb{F}_2^{n\times n} \rightarrow \mathbb{F}_2$ be an algorithm running in time~$T(n)$ such that for every sufficiently large~$n$, 
    \[
    \Pr_{A\sim \mathbb{F}_2^{n\times n}} \Big[
     \mathcal{M}\left(A\right) = det(A) 
    \Big] \geq \frac{2+q_2}{3} + \varepsilon.
    \]
    Then, we can construct an algorithm that for \textbf{any} matrix $A\in \mathbb{F}_2^{n\times n}$ computes $det(A)$ with high probability, in time~$\tilde{O}\left(T(n)\right)$.
\end{lemma}
\begin{proof}
    Initialize a counter~$C=0$.
    For~$t=\Theta\left(\log \left(n\right) /\varepsilon^2\right)$ times we repeat the following: Draw a matrix~$R$ from our trapdoored family, and query the algorithm on~$\mathcal{M}(R)$ and on~$\mathcal{M}(RA)$;
    If~$\mathcal{M}(R)=\mathcal{M}(RA)$ then increment~$C$.
    Finally, return~$1$ if and only if~$C<\frac{2}{3}(1-q_2)\cdot t$ and otherwise return~$0$.

    If~$A$ is invertible, that is~$det(A)=1$, then both~$R$ and~$RA$ are distributed uniformly and also~$det(RA)=det(R)det(A)=det(R)$. In particular, the probability that we increment the counter in any iteration is bounded by~$2\cdot\left(1-\left(\frac{2+q_2}{3} + \varepsilon\right)\right) = \frac{2}{3}(1-q_2)-2\varepsilon$. Thus, with high probability, we return~$1$.

    If~$A$ is singular, then pick a basis of~$\mathbb{F}_2^n$ in which the last vector is in~$\text{Ker}(A)$. We note that both the distribution of uniform matrices and determinants are invariant to this change of basis. 
    When drawing a uniformly random~$R$, the probability that the first~$(n-1)$ vectors of~$R$ will be linearly independent is at least~$2(1-q_2)$. Denote this event by~$B$.
    Conditioned on~$B$, we observe that~$det(R)$ is uniformly distributed in~$\mathbb{F}_2$ as the last vector of~$R$ is either in or out of the span of the first~$n-1$ vectors with equal probability.
    Furthermore, conditioned on~$B$ the matrix~$RA$ is independent of~$det(R)$ as~$RA$ is fully determined by the first~$n-1$ vectors of~$R$ (as the last one is in the kernel of~$A$).
    In particular, conditioned on~$B$ and on the event that~$\mathcal{M}(R)=det(R)$, we have that~$\mathcal{M}(R)\neq\mathcal{M}(RA)$ with probability~$\frac{1}{2}$ regardless of what~$\mathcal{M}(RA)$ is. The probability of incrementing~$C$ is thus at least~$2(1-q_2)\cdot \frac{1}{2} - \left(1 - \left(\frac{2+q_2}{3} + \varepsilon\right)\right) = \frac{2}{3}(1-q_2) + \varepsilon$.
    Thus, with high probability, we return~$0$.
\end{proof}

We {remark} that as~$p\rightarrow \infty$, the correctness probability we require for the average-case algorithm approaches~$\frac{3}{4}$.
As the algorithm which simply always returns~$0$ succeeds with probability~$q_p\pm o(1)$, we cannot obtain a reduction that uses any positive correctness probability. 
Furthermore, consider an algorithm~$\mathcal{M}(A)$ that returns either~$det(A)$ or~$-det(A)$ with equal probability. The algorithm succeeds with a probability strictly larger than~$\frac{1}{2}$ yet it is unclear if there exists a reduction from computing the determinant to computations of such unsigned determinants. In particular, most operations involving the determinant are multiplicative and hence are not immediately useful to recover the sign. 
If we could construct a family of trapdoored matrices in which we can quickly compute the determinant of the sampled matrix, then similar arguments to those we used would require correctness probability approaching~$\frac{1}{2}$ as~$p\rightarrow \infty$.
It is thus reasonable to conjecture that~$\frac{1}{2}$ is the correct threshold for when such reductions are possible over a large field. 

\section{Summary, Discussion and Open Problems}
The main contribution of this work is conceptual: we propose accelerating classical algorithms by replacing a truly random object with a carefully structured pseudorandom object that comes with a (hidden) trapdoor that enables an equivalent but more efficient computation. We believe this idea opens a much broader direction of inquiry, and to that end, we raise the question of whether similar constructions can be applied beyond random matrices to other classes of random objects.

This suggests a general framework:  we are interested in meaningful sets of actions $\mathcal{G}$ that act on a set of objects $\mathcal{X}$. In the exemplary scenarios explored in this paper, $\mathcal{G}$ is the set of all linear transformations (represented as matrices) and $\mathcal{X}$ is either the set of all vectors, or the set of all matrices themselves.  We ask: are there other interesting  
{\em distributions over actions} $\mathcal{D}(\mathcal{G})$ for which the following holds?
\begin{itemize}
\item Applying a random action $g \sim \mathcal{D}(\mathcal{G})$ (chosen from the prescribed distribution) to an object $x \in \mathcal{X}$ is computationally expensive.
\item There exists a {\em pseudorandom} distribution $\tilde{\mathcal{D}}$ of actions, indistinguishable from the original, for which performing the same action is {\em significantly faster}, given a trapdoor $t_g$ for $g \sim \mathcal{D}(\mathcal{G})$.
\end{itemize}

Beyond matrix-vector and matrix-matrix multiplication, other potential examples in the matrix setting include operations such as repeated exponentiation, which can also be viewed as taking multiple steps in a random walk. More broadly, other random structures that may admit similar trapdoor constructions include random graphs, random sets, and random polynomials. We leave the exploration of these objects and their potential applications to future work.

We also leave open the problem of further supporting or refuting Conjecture~\ref{conj:kac}. This could involve proving that the conjecture follows from standard cryptographic assumptions, or alternatively, demonstrating that plausible classes of attack strategies, e.g. attacks implementable via low-degree polynomials, are provably ineffective.

Finally, several problems remain open in connection to the (conditional) worst-case to average-case reductions for linear algebraic operations. In particular, can errors be corrected for the problems in addition to matrix multiplication? What is the lowest success probability from which an average-case algorithm for the determinant can be used for worst-case inputs? And for which other operations such reductions can be obtained?

\ifsub
\else
\paragraph{Acknowledgements.} The authors would like to thank Mark Braverman and Stephen Newman for sending us a draft of their work~\cite{DBLP:journals/corr/abs-2502-13060}.
\fi 

\bibliographystyle{alpha}
\bibliography{refs}

\end{document}